\documentclass[conference]{IEEEtran}
\IEEEoverridecommandlockouts
\usepackage{cite}
\usepackage{amsmath,amssymb,amsfonts}
\usepackage{algorithmic}
\usepackage{graphicx}
\usepackage{textcomp}
\usepackage{xcolor}
\usepackage{authblk}
\usepackage{url}
\usepackage{times}
\usepackage{latexsym}
\usepackage{amsmath}
\usepackage{xr}
\DeclareMathOperator*{\argmax}{arg\,max}

\usepackage[T1]{fontenc}
\usepackage[utf8]{inputenc}

\usepackage{microtype}

\usepackage{inconsolata}

\usepackage{comment}

\usepackage{color}

\usepackage{multirow}
\usepackage{amssymb}
\usepackage{caption}
\usepackage{subcaption}

\newcommand{\CG}{\textsc{CodeGen}}
\newcommand{\WIS}{\textsc{Wisdom}}
\newcommand{\CGNL}{\textsc{CodeGen-NL}}
\newcommand{\CGMU}{\textsc{CodeGen-Multi}}
\newcommand{\CGMO}{\textsc{CodeGen-Mono}}
\newcommand{\WA}{\textsc{Wisdom-Ansible}}
\newcommand{\WY}{\textsc{Wisdom-Yaml}}
\newcommand{\WAM}[1]{\textsc{Wisdom-Ansible-Multi#1}}
\newcommand{\WYM}{\textsc{Wisdom-Yaml-Multi}}
\newcommand{\CODEXEDIT}{Codex-Davinci-002}
\newcommand{\CODEXCOMPLETION}{Codex-Davinci-002}

\newcommand{\NLtoPB}{\textsc{NL$\rightarrow$PB}}
\newcommand{\PBNLtoT}{\textsc{PB+NL$\rightarrow$T}}
\newcommand{\NLtoT}{\textsc{NL$\rightarrow$T}}
\newcommand{\TNLtoT}{\textsc{T+NL$\rightarrow$T}}

\usepackage{array}
\newcolumntype{C}[1]{>{\centering\arraybackslash}p{#1}}

\definecolor{DarkGreen}{RGB}{0,100,0}
\definecolor{DodgerBlue4}{RGB}{16,78,139}
\newcommand\YAMLcolonstyle{\color{black}\mdseries}
\newcommand\YAMLkeystyle{\color{DarkGreen}\bfseries}
\newcommand\YAMLvaluestyle{\color{DodgerBlue4}\mdseries}
\usepackage[hidelinks]{hyperref}
\usepackage{listings}
\lstdefinelanguage{yaml}{
     keywordstyle=\color{darkgray}\bfseries,
     basicstyle=\YAMLkeystyle,
     sensitive=false,
     comment=[l]{\#},
     morecomment=[s]{/*}{*/},
     commentstyle=\color{purple}\ttfamily,
     stringstyle=\YAMLvaluestyle\ttfamily,
     moredelim=[l][\color{orange}]{\&},
     moredelim=[l][\color{magenta}]{*},
     moredelim=**[il][\YAMLcolonstyle{:}\YAMLvaluestyle]{:},   % switch to value style at :
     morestring=[b]',
     morestring=[b]",
     literate =  
                {>}{{\textcolor{red}\textgreater}}1     
                {|}{{\textcolor{red}\textbar}}1
                {-}{{\color{black}\mdseries{-}\ }}1,
      }

\def\BibTeX{{\rm B\kern-.05em{\sc i\kern-.025em b}\kern-.08em
    T\kern-.1667em\lower.7ex\hbox{E}\kern-.125emX}}
\begin{document}

\title{Automated Code generation for Information Technology Tasks in YAML through Large Language Models}

\author[1]{Saurabh Pujar$^*$}
\author[1]{Luca Buratti$^*$}
\author[1]{Xiaojie Guo$^*$}
\author[1]{Nicolas Dupuis$^*$}
\author[1]{Burn Lewis$^*$}
\author[1]{Sahil Suneja$^*$}
\author[1]{\\ Atin Sood$^*$}
\author[2]{Ganesh Nalawade$^*$}
\author[2]{Matthew Jones}
\author[1]{Alessandro Morari}
\author[1]{Ruchir Puri}
\affil[1]{IBM Research}
\affil[2]{Red Hat}

\maketitle

\begin{abstract}
The recent improvement in code generation capabilities due to the use of large language models has mainly benefited general purpose programming languages. Domain specific languages, such as the ones used for IT Automation, have received far less attention, despite involving many active developers and being an essential component of modern cloud platforms.
This work focuses on the generation of Ansible-YAML, a widely used markup language for IT Automation.
We present Ansible Wisdom, a natural-language to Ansible-YAML code generation tool, aimed at improving IT automation productivity. Ansible Wisdom is a transformer-based model, extended by training with a new dataset containing Ansible-YAML. We also develop two novel performance metrics for YAML and Ansible to capture the specific characteristics of this domain. 
Results show that Ansible Wisdom can accurately generate Ansible script from natural language prompts with performance comparable or better than existing state of the art code generation models.
In few-shot settings we asses the impact of training with Ansible, YAML data and compare with different baselines including \CODEXEDIT.
We also show that after finetuning, our Ansible specific model (BLEU: 66.67) can outperform a much larger \CODEXEDIT~(BLEU: 50.4) model, which was evaluated in few shot settings.
%\footnote{Invited Paper}
\end{abstract}
\def\thefootnote{*}\footnotetext{Equal contribution. Correspondence: saurabh.pujar@ibm.com}
%\def\thefootnote{}\footnotetext{Correspondence: saurabh.pujar@ibm.com}
%\begin{IEEEkeywords}
\begin{IEEEkeywords}
Generative Model, Ansible, Code Generation
\end{IEEEkeywords}

\section{Introduction}
\label{intro}

In the recent years, Large Language models (LLMs) have demonstrated considerable content-generation capabilities in multiple domains, including natural language, vision, video and audio processing \cite{openai2023gpt4}. 
More recently, LLMs have been applied to the Software Engineering field, with the objective of improving aspects such as programmer’s productivity~\cite{jain2022jigsaw} and software security~\cite{buratti2020exploring}.
In the space of general-purpose programming languages, a growing amount of research is exploiting the capabilities of large language models to perform code generation, clone detection, code repair and other tasks~\cite{puri2021codenet, greengard2023}. This is fueling a new generation of coding assistants’ products, able to speed up developer work, leveraging the availability of open-source code bases to train the models. While the field of AI-assisted coding assistants is at its infancy, the potential impact on the field of Software Engineering cannot be underestimated. 
The application of these techniques to IT domain specific languages like YAML, however, has received less attention, despite their importance to a wide array of fields. 

In this work, we explore the application of LLMs to the important area of IT automation. IT automation replaces manual IT admin work with automated execution of domain specific scripts. This approach dramatically improves cloud infrastructure security, cost-efficiency, reliability and scalability. 
YAML files are often used to define and configure key aspects of IT infrastructure.
Ansible is one of the most widely used applications for IT automation which uses YAML-based configurations, and thousands of companies rely on this technology to manage their IT infrastructure.  
% Include Ansible example/figure
While easier to write than general purpose programming languages such as Java and C++, Ansible-YAML requires considerable expertise to be used proficiently. For many companies, speeding up Ansible adoption would mean faster digital transformation towards a safer, cost-efficient approach for IT infrastructure management. 
% what is correct? ansible code or ansible script or ansible yaml?
This paper investigates the use of LLMs to generate Ansible-YAML code, with the objective of building an AI assistant for Ansible-YAML users and improving their productivity. We propose the use of transformer-based models for the task of Ansible-YAML code generation given a natural language prompt. We start from training four versions of large domain-specific pre-trained decoder-based model, by learning from large amount of YAML and Ansible-YAML data in general. We then perform fine-tuning for the downstream Natural Language to Ansible-YAML generation task. 
This is the first work looking at LLMs for YAML in general and for Ansible-YAML in particular.
The contributions of this work are the following:
\begin{itemize}
    \item We explore the implications of applying code generation to Ansible-YAML and provide a formal definition of the problem.
    \item We build the YAML and an Anisible-YAML dataset for both pretraining and finetuning tasks in the code generation.
    \item We theoretically re-formalize the Ansible-YAML generation problem into code completion with novel prompt, by utilizing the unique features of YAML data and practically trained a series of transformer-based models, which show much superiority.
    \item We propose two novel evaluation metrics specially designed for Ansible-YAML, compare our models against the latest LLMs, and highlight their limitations.
\end{itemize}

\section{Related Work}

% \subsection{Ansible YAML}
% \todo{Ale: finish this}
% ~\cite{borovits_findici_2022}
% ~\cite{quattrocchi_predictive_2022}
% ~\cite{dalla_palma_ansiblemetrics_2020}
% ~\cite{opdebeeck_practice_2021}

\subsection{Pre-trained Language Models for Code}

Most recently, language models have fueled progress towards the longstanding challenge of source code synthesis~\cite{chen2021evaluating, touvron2023llama}, which excel at  downstream tasks such as code completion, code generation and code summarization.

%For example, CodeBERT~\cite{feng2020codebert} trained the BERT objective on doc-strings paired with functions for code search. One of the largest models, Codex~\cite{chen2021evaluating}, a decoder-based language model, has been deployed in the real-world production tool GitHub Copilot. 
According to Xu et al.~\cite{xu2022systematic}, pre-training methods for source code modeling fall into three categories: (i) The first category is based on \textit{left-to-right language models}, namely, auto-regressive decoder-based models. These models predict the probability of a token given the previous tokens. For example, CodeGPT~\cite{lu2021codexglue}, CodeParrot~\cite{tunstall2022natural}, Codex~\cite{chen2021evaluating}, AlphaCode~\cite{li2022competition} and CodeGen~\cite{nijkamp2022conversational} all follow into this line, which are highly useful for code generation and completion tasks. (ii) The second category is based on \textit{masked language models}, which can make use of the bi-directional information to learn whole sentence representations, such as CodeBERT~\cite{feng2020codebert} and CuBERT~\cite{kanade2020learning}. This line of pre-trained models perform well for the code classification and detection tasks. (iii) The third category of models is based on \textit{encoder-decoder models} that incorporate pre-training objectives such as masked span prediction and denoising sequence reconstruction. CodeT5~\cite{wang2021codet5}, PLBART~\cite{ahmad2021unified}, and PolyCoder~\cite{xu2022systematic} fall into the third category and perform well in sequence-to-sequence downstream tasks such as code commenting and code-to-code translation.

Among these models, CodeGen has been trained on The Pile~\cite{gao2020pile} and on data from Google BigQuery, hence it has been exposed to natural language, code and some YAML data.

\subsection{Code Generation}
Source code generation (or program synthesis) can be defined as the generation of a program or code snippet, starting from a natural language specification. Traditional methods use a probabilistic context free grammar (PCFG) to generate the abstract syntax tree (AST) of the source code~\cite{chen2021evaluating, yin2017syntactic}. Yin et al.~\cite{yin2017syntactic} proposed a neural model in combination with a transition system to generate abstract syntax trees. 

With the recent development of large scale language models, large scale Transformers have also been applied to this problem. Transformers typically treat code as text. Feng et al.~\cite{feng2020codebert} proposed the use of a masked language model with bi-modal text from the CodeSearchNet challenge~\cite{husain2019codesearchnet}.
Two works~\cite{austin2021program, chen2021evaluating} propose the use of a decoder language model trained on large amounts of source code and web data. Furthermore, Xu et al.~\cite{xu2022systematic} perform a systematic evaluation of these models, finding that the presence of natural language in the training corpus helps with general code-language modeling.

Many transformer models have been developed for software engineering tasks which focus on specific programming languages like Python~\cite{kanade2020learning} and C~\cite{buratti2020exploring}. 
Multi-lingual LLMs like PLBART~\cite{ahmad2021unified}, \CG{}~\cite{nijkamp2022conversational} and Codex~\cite{chen2021evaluating} are trained on multiple programming languages but most of the data is comprised of commonly used languages like C, C++, Java, Python etc.

Widely used domain specific languages like YAML have received far less attention. 
Tools such as Ansible, OpenShift and many others rely on YAML for managing their configuration files.
State of the art models such as Codex~\cite{chen2021evaluating} or \CG{}~\cite{nijkamp2022conversational}, are primarily evaluated on general purpose programming languages. While Codex or \CG{} could be used to generate YAML, including Ansible-YAML, due to their very large and heterogeneous training datasets, we did not find any work evaluating this capability. To the best of our knowledge, this is the first work addressing the problem of YAML code generation using a large language model.

\section{Background}
The Red Hat Ansible Automation Platform is an open-source \cite{Ansible, githubAnsible} IT automation system. It handles configuration management, application deployment, cloud provisioning, ad-hoc task execution, network automation, and multi-node orchestration. Ansible makes complex changes like zero-downtime rolling updates with load balancers simpler. A system running Ansible will have a control node and one or more managed nodes. The control node is where Ansible is executed, while the managed nodes are the devices being automated, for example, Linux and Windows server machines. 

An \textit{Ansible Playbook}  (or playbook) is a YAML file that describes a set of \textit{Ansible Tasks} (or tasks) to be performed by Ansible on the managed node. The playbook defines the desired state of the managed nodes, and the tasks specify the steps to bring the nodes to that desired state. For example, a playbook might define a set of tasks to install and configure a particular application, or to set up a particular system configuration.

Playbooks are organized into a series of plays, which are executed in order. Each play specifies a group of managed nodes and a set of tasks to be performed on those nodes. Playbooks can also include variables and conditional statements, which allow for more flexible and dynamic execution. This makes it easy to define complex configurations and deploy them consistently across a fleet of servers. Fig.~\ref{fig:ansible_example} shows an example of an Ansible playbook.
\begin{figure}[h]
\setlength{\belowcaptionskip}{-6pt}
\lstset{
  basicstyle=\tiny,	
  xleftmargin=.1\textwidth, xrightmargin=.1\textwidth
}
\begin{lstlisting}[language=yaml,columns=flexible]
---
- hosts: servers
  tasks:
    - name: Install SSH server
      ansible.builtin.apt:
        name: openssh-server
        state: present
    - name: Start SSH server
      ansible.builtin.service:
        name: ssh
        state: started
\end{lstlisting}
\caption{Example of an Ansible playbook}
\label{fig:ansible_example}
\end{figure}
The playbook in Fig.~\ref{fig:ansible_example} consists of a single play that targets all managed nodes in the server group. The play includes two tasks: the task named ``Install SSH server'' uses the \textit{ansible.builtin.apt} module to install the \textit{openssh-server} package, and the task named ``Start SSH server'' uses the \textit{ansible.builtin.service} module to start the ssh service. The ``name'' field of each task can be customized by users to describe the intention of the task. 

\section{Methodology}

\label{sec:method}
    \subsection{Problem formulation}
    \label{subsec:problem}
    We define the task \textit{Ansible-YAML Generation} as follows:
    %Given a task description $X=[x_1, x_2,...,x_N]$ including both natural language (NL) as prompt 
    %BLL restructured this line
    given a task description that includes both natural language (NL) as prompt $X$
    %=[x_1, x_2,...,x_N]$,
    and Ansible YAML as the context script $C$,
    %=[c_1, c_2,...,c_M]$, 
    generate an Ansible task or playbook snippet $Y$ based on the intent of $X$ and $C$. Both $X$ and $C$ are represented as a sequence of tokens.  %where $N$ is the length of the input sequence. 
    The snippet code $Y$ is also formalized as an Ansible Language sequence (AL).
    %$Y=[y_1, y_2,...,y_L]$.
    %BLL shorten sentence ... lengths are explicitly defined in the figure.
    %, where $L$ is the length of sequence. \\
    We also define a \textit{Probabilistic generative model} to model the distribution of an Ansible snippet $Y$ given $X$ and $C$ as $p(Y|X,C)$. The best-possible Ansible task snippet is then given by
    \begin{equation}
        \hat{y}=\argmax p(Y|X, C).
        \label{eq:1}
    \end{equation}

\subsection{Dataset Construction}

We curated a YAML dataset from multiple data sources, including GitHub, Google BigQuery\footnote{A publicly available dataset published by Google, https://cloud.google.com/bigquery}, GitLab and Ansible Galaxy~\cite{AnsibleGalaxy}.
We use data extraction logic specific to each data source, while querying their respective API endpoints to extract YAML files and relevant associated metadata. For Google BigQuery, we downloaded every file with a valid YAML extension (`.yml', `.yaml'). For GitHub and GitLab, we considered every repository containing ``Ansible'' either in the name or the description. 
We de-duplicated the dataset using a simple exact match criterion.
In addition to generic YAML, our dataset contains ansible-specific YAML, appropriately tagged so as to preserve the interplay between Ansible roles, collections, tasks and playbooks.

Our curated dataset contains $\sim$1.1M Ansible task and playbook YAMLs, and $\sim$2.2M other generic YAML files.
Table~\ref{tab:table-dataset-size} summarizes our data sources, file count, type of YAML, and whether we use the data for pre-training (PT) of fine-tuning (FT). 
%also distinguishes between data used for pre-training %vs. fine-tuning.

\begin{table}
\centering
\begin{tabular}{p{0.12\textwidth}|*{3}{C{0.05\textwidth}}}
\hline
\textbf{Source} & \textbf{File Count} & \textbf{YAML Type}&
\textbf{Usage}\\
\hline
\hline
%Galaxy & 226K & Ansible & FT \\
%BL-Galaxy #s after deduplication and discarding playbooks without tasks
Galaxy & 112K & Ansible & FT \\
GitLab & 64K & Ansible & PT\\
GitHub + GBQ & 1.1M & Ansible & PT\\
GitHub + GBQ & 2.2M & Generic & PT\\
\hline
\end{tabular}
\caption{Extracted file count per data source. The data is used for pre-training (PT) or fine-tuning (FT) of Wisdom models.}
\label{tab:table-dataset-size}
\end{table}

%SSP: Expect Ansible Playbook, task, role description here.
\subsection{Pre-training}
Our pre-trained models are implemented with the same architecture as \CG{}, a decoder-based model released by SalesForce~\cite{nijkamp2022conversational}.
\CG{} has been pre-trained on several datasets: 
(1) the Pile~\cite{black2022gpt}, around 350 billion tokens of natural language and 31 billion tokens of code; 
(2) BigQuery,  around 119 billion tokens of code in 6 programming languages;
(3) BigPython, around 71 billion tokens of Python code.
%Our baseline is \CG{} trained on the Pile + BigQuery, but we also tested \CG{} trained on the Pile alone.
\CG{} has seen a large amount of natural language, but only a limited number of Ansible-YAML. For example,
the Pile only includes around 25K Ansible-YAML and 600K 
generic YAML files.
To improve the pre-trained model understanding of the semantics and syntax of YAML, we build \WAM{} and \WYM{}, which are trained from the \CG{} checkpoint with 
a dataset that contains only Ansible-YAML files and a dataset that contains Ansible-YAML and generic YAML files, respectively (see table~\ref{tab:table-dataset-size} for detail). 
The Ansible-YAML and generic YAML files account for about 1.1 billion training tokens in total. In addition, to exploring the effectiveness of using \CG{} checkpoint as initialization, we propose \WA{} and \WY{}, which are trained from scratch with the above mentioned two datasets.

Wisdom is designed to assist Ansible programmers in real-time and latency is therefore a critical parameter to consider.
Which is why we choose a reasonably-sized model with 
a high token-per-second throughput rather than a very large model with a low throughput.
We tested the architecture of \CG{} 350M and \CG{} 2.7B.
We benchmarked the generation throughput on single GPU for both models and found that the 350M model was 
$\sim 1.9\times$ faster than than the 2.7B.

Our training code is based on the Huggingface Transformers library~\cite{wolf-etal-2020-transformers} that provided the \CG{} checkpoints and tokenizers. 
We trained the model using our YAML dataset for 9 epochs using 16 A100 GPUs with $80~$GB of memory.
To speed up the training we used bf16 data type.
Effective batch size was $32$ and learning rate was $5\times10^{-5}$ with a linear decreasing schedule.
During pre-training, YAML files were packed to fill up a context window of $1024$, and we used
a special separator token to separate the files.

\subsection{Fine-tuning}
\label{subsec:data_ft}

% Moved the basic validation steps to the middle of the dataset section
%\subsubsection{Validation}  % \todo{Talk about Ansible syntax check [Burn]}
%The Galaxy data contains many type of Ansible files, but for our fine-tuning data we extracted only playbooks containing tasks, and lists of tasks from roles.
%We checked their syntax using PyYAML (\url{https://pyyaml.org}) and standardized their formatting into the style recommended by the Ansible team.  
% Could discuss future steps elsewhere
%Additional modifications under consideration are to convert any tasks using the deprecated \textit{key=value} syntax for module parameters, and to discard any that don't satisfy the Ansible playbook or task schemas.  
%Although the schemas reject some old-style Ansible tasks that are still valid, this would help the model generate good-quality preferred Ansible.

\subsubsection{Dataset}
\label{sssec:galaxy_dataset}
%The \CG{} model needs to be fine-tuned for the \textit{Natural Language to Ansible generation} task as the user intent is expressed in natural language and this may not always be available in the Ansible data used for pre-training.
We used the Ansible Galaxy data to fine-tune the pretrained models mentioned above on the \textit{Ansible-YAML generation} downstream tasks,
as this dataset is a collection of good quality files created and vetted by the Ansible community.
Galaxy contains many type of Ansible files, but we extracted only playbooks containing tasks, and lists of tasks from roles.
We checked for valid YAML and correct playbook or task syntax using PyYAML (\url{https://pyyaml.org}), and standardized the formatting to match the style recommended by the Ansible team.  
The Galaxy data files were randomly split into train (80\%), validation (10\%) and test (10\%) sets. 
%To maintain good quality of the fine-tuning datasets, we collect data from Ansible Galaxy dataset, which correct and documented. 
Exact match deduplication is performed at both the file and sample level across all splits. 
%More details about data collection can be found in the Appendix.

%The input can either include the Natural Language (NL) intent of the user or include both of the NL intent and the context script.
%To generate a full playbook, the input context will only be the Natural Language intent.
%To generate a task, the input context can contain, apart from natural language intent, the preceding playbook, the preceding role or nothing.

\subsubsection{Generation Types}
\label{subsec:generationtypes}
As described in Section~\ref{subsec:problem}, the goal is to generate two kinds of Ansible output, either a full playbook (PB) or a task (T) given the natural languages (NL) requirement.
The task (T) can be either part of a playbook, or part of a role. Thus, we can have 4 types of input-output combinations in the fine-tuning dataset.

\begin{itemize}
    \item \textbf{\NLtoPB{}}: The context is empty, so the only input is the natural language prompt.
We have limited the expected output playbooks  to examples containing only 1 or 2 tasks. 
This forms the vast majority of playbooks. 
Playbooks containing more than 2 tasks are used to generate the next type of samples.
%multiple type 3 samples.
\item \textbf{\PBNLtoT{}}: 
%Similar to type 4, but 
The model is expected to predict the next task in a Playbook, and
the context is a playbook with at least 1 task.
\item \textbf{\NLtoT{}}: 
%In this case as well, the input context empty, so the only input is the natural language prompt.
The context is empty and
the model is expected to generated only 1 task, which is the first task of a role.
\item \textbf{\TNLtoT{}}: The model is expected to predict the next task in a role based on the natural language prompt,
where the context is the previous tasks.
%are passed to the model as context, but if the context is too big the oldest tasks are pruned.

\end{itemize}

\begin{figure*}[!t]
\begin{center}
\begin{subfigure}[b]{0.4\linewidth}
\begin{lstlisting}[language=yaml,columns=flexible,
basicstyle=\footnotesize,numbers=left,numberstyle=\ttfamily]
---
# Generating a task from NL prompt (L18) 
# using a playbook as context (L1-L17)
# model expected output in (L19-L20) 
- name: Network Setup Playbook
  connection: ansible.netcommon.network_cli
  gather_facts: false
  hosts: all
  tasks:
    - name: Get config for VyOS devices
      vyos.vyos.vyos_facts:
        gather_subset: all
    - name: Update the hostname
      vyos.vyos.vyos_config:
        backup: yes
        lines:
          - set system host-name vyos-changed
    - name: Get changed config for VyOS devices
      vyos.vyos.vyos_facts:
        gather_subset: all
\end{lstlisting} 
%\vspace{-2mm}
\caption{\PBNLtoT{}}
\label{fig:original_code}
\end{subfigure}
\qquad
\begin{subfigure}[b]{0.4\linewidth}
\begin{lstlisting}[language=yaml,columns=flexible,
basicstyle=\footnotesize,numbers=left,numberstyle=\ttfamily]
---
# Generating a playbook from NL prompt (L5)
# without  context
# model expected output in (L6-L17) 
- name: Network Setup Playbook
  connection: ansible.netcommon.network_cli
  gather_facts: false
  hosts: all
  tasks:
    - name: Get config for VyOS devices
      vyos.vyos.vyos_facts:
        gather_subset: all
    - name: Update the hostname
      vyos.vyos.vyos_config:
        backup: yes
        lines:
          - set system host-name vyos-changed
\end{lstlisting} 
\vspace{5mm}
\caption{\NLtoPB{}}
\label{fig:clone_deviant}
\end{subfigure}
\qquad
\begin{subfigure}[b]{0.4\linewidth}
\begin{lstlisting}[language=yaml,columns=flexible,
basicstyle=\footnotesize, numbers=left,numberstyle=\ttfamily]
---
# Generating a task from NL prompt (L9) 
# using task(s) as context (L1-L8)
# model expected output in (L10-L12)
- name: Ensure apache is at the latest version
  ansible.builtin.yum:
      name: httpd
      state: latest
- name: Write the apache config file
  ansible.builtin.template:
      src: /srv/httpd.j2
      dest: /etc/httpd.conf
\end{lstlisting}
%\vspace{-2mm}
\caption{\TNLtoT{}}
\label{fig:synthetic_clone}
\end{subfigure}
%\vspace{-3mm}
\qquad
\begin{subfigure}[b]{0.4\linewidth}
\begin{lstlisting}[language=yaml,columns=flexible,
basicstyle=\footnotesize,numbers=left,numberstyle=\ttfamily]
---
# Generating a task from NL prompt (L5) 
# without context
# model expected output in (L6-L8)
- name: Ensure apache is at the latest version
  ansible.builtin.yum:
      name: httpd
      state: latest
\end{lstlisting} 
\vspace{8mm}
\caption{\NLtoT{}}
\label{fig:gentype}
\end{subfigure}
\caption{Ansible generation types defined in~\ref{subsec:generationtypes}.
Each snippet of code includes a comment (in red) highlighting the NL prompt, the context provided to the model as well as the expected output. The comments are used here only for illustration purpose and are not provided to the model during training or inference.}
\end{center}
\end{figure*}

\subsubsection{Input Prompt Formulation}
\label{sec:prompt_formulation}
A helpful feature of the Ansible language is that each 
Playbook or Task frequently contains a ``name'' field, whose value is the natural language description of the goal of playbook or task, as shown in Fig.~\ref{fig:ansible_example}. 
% BLL just reversed the order as the name is usually the 1st line in the task
Thus the target output can be represented as $Y=\{Y_{NL},Y_{AL}\}$, where $Y_{NL}$ refers to the ``name'' line in the Ansible script, and $Y_{AL}$ refers to the remainder of the script.
%Thus, the target output can be represented as $Y=\{Y_{AL},Y_{NL}\}$, where $Y_{AL}$ refers to the ansible script excluding the `name' field, and $Y_{NL}$ refers to the `name' field. 
In addition, $Y_{NL}$ is exactly the same as the NL sequence $X$ in the original problem formulation in Section~\ref{subsec:problem}.

Thus, to take advantage of this feature and to make it accommodate best the pre-trained decoder-based model, we re-formalize the text-to-code generation problem in Section~\ref{subsec:problem} into a code completion problem. Specifically, Eq.\ref{eq:1} can be formalized as 
% BLL should we reverse the order of the 2 Y variables here?
    \begin{align}\nonumber
        %&= p(Y_{AL}|Y_{NL}, C) 
        %\hat{y}=&\argmax p(Y_{AL},Y_{NL}|X, C).\\\nonumber
        %&=p(Y_{AL}|X, C)p(Y_{NL}|X,C) \\\nonumber
        %&=p(Y_{AL}|Y_{NL}, C)p(X|X,C) \\
        %&= p(Y_{AL}|Y_{NL}, C) 
        \hat{y}=&\argmax p(Y_{NL},Y_{AL}|X,C).\\\nonumber
        &=p(Y_{NL}|X,C) p(Y_{AL}|X,C) \\\nonumber
        &=p(X|X,C) p(Y_{AL}|Y_{NL},C) \\
        &=p(Y_{AL}|Y_{NL}, C) 
        \label{eq:2}
    \end{align}
considering that $Y_{AL}$ and $Y_{NL}$ are conditionally independent given $X$ and $C$ and $X=Y_{NL}$. Thus, we can use the value of the ``name'' line $Y_{NL}$ as the prompt.
% For Type 1 samples,   BLL trying to avoid the sample-type numbers 
When the output is a playbook,
we combine the values of ``name'' fields of the playbook and its tasks to create the prompt. We have experimentally validated that this re-formalization can largely improve the overall performance regarding all kinds of metrics. The results are provided in the following section. \\

\subsubsection{Training}
We fine-tuned pre-trained models using our Galaxy dataset for 8 epochs.
The effective batch size was $32$ and the learning rate was $5\times10^{-5}$ with a cosine decreasing schedule.
We used the \textsc{Bleu} score on the validation set to determine the best checkpoint.
 
\subsection{Demo/Plugin}
We expose a GRPC and REST API based interface to model predictions so that inference can be called out using GRPC and REST clients. 
We wrote a custom Visual Studio Code plugin that is enabled for ansible files and gets triggered when the user hits a binding key. This triggers a call to the API to carry out the prediction which is then formatted and pasted back on to the editor.
In our current setup, when a user writes the prompt for the task, example ``\textit{- name: install nginx on RHEL}'', and hits enter, we invoke the API to carry out the prediction and then take the results and paste it back on the editor. The user can either hit tab and accept the suggestion, or escape key to reject the suggestion.
In future implementations, we plan to improve user experience in terms of quality of recommendation by leveraging additional information in workspace of the editor, as well as improving latency by using techniques like caching. 
%Details about splits

%\subsubsection{Ansible Task and Playbook Description}
%1. Ansible Galaxy 2. Task/playbook 3. name field and prompt
%\subsubsection{Dataset construction}
%1. Breaking of files 2. Filters 3. splits
%\subsubsection{Description}
%1. Fields 2. Sample count 3. others 4. Examples    
\section{Experiments}

\subsection{Evaluation Metrics}
\label{subsec:eval_metrics}

Since the generated ansible task or a small playbook always has high dependency on external resources, it is not practical to evaluate the correctness of a task by executing it.
For example it would be impractical to evaluate a task that installs a package on a number of remote hosts by executing it with Ansible and checking that the result is as expected. Thus, our evaluation metrics are based on the similarity between the generated ansible tasks or playbooks and the ground-truth.

For the experiments described in this paper, 4 comparison metrics are used: \textbf{Exact Match, \textsc{Bleu}}~\cite{ibm2001bleu,lin2004orange}\footnote{Since the sequences of tokens in an Ansible YAML file are important, while some reordering is permitted, the \textsc{Bleu} score's basis on n-gram coverage suggests it could be a useful metric.}, \textbf{Ansible Aware}, and \textbf{Schema Correct}. Among these, \textbf{Ansible Aware} and \textbf{Schema Correct} are two novel metrics designed specially for the Ansible tasks or playbooks.\\

\noindent\textbf{Ansible Aware}: Ideally a metric should reflect the user's view of the result, e.g. how many changes must be made to correct it.
%One approach is to compute some sort of graph edit distance, so we have used DeepDiff~\cite{deepdiff} to compute the difference between the two Python objects created from the YAML files.
The purpose of the Ansible-aware metric is to use knowledge of the Ansible YAML syntax to compare the modules, keywords and parameters that comprise an Ansible task or playbook.
%For the experiments described in this paper we have computed these 3 comparison metrics: Exact Match, \textsc{Bleu}, Ansible Aware.
%\smallskip

%\noindent
%\textbf{Deep Diff (DD)}
%  \subsubsection{DeepDiff (DD)}
%The DeepDiff function in the Python package (\url{https://pypi.org/project/deepdiff}) computes the difference between two python objects and produces a detailed list of the changes.
%We create a 0-1 score from the number of changes and deletions, normalized by the number of keys in the file.
%Since we want the score to reflect a correction cost we decided to ignored insertions (they are easier to fix than deletions.)
%\smallskip

%\noindent
%\textbf{Ansible Aware (AA)}
 % \subsubsection{Ansible-Aware (AA)}
Since an Ansible task or playbook is a mapping (dictionary) the order of the key-value pairs is not significant --- the usual key order for a task is: \textit{name, module, keyword(s)}.

The \textit{``name''} is optional, its value is a natural language description of the task.
The module key identifies the operation to be performed while its value is a dict holding the module's parameters.
The optional keywords define conditions that influence the execution of the task,
e.g. environment, elevated privileges, remote userid, error handling, conditionals, loops.
The keyword values may be scalars, lists, or dicts.
The score of a task is computed from the average of the scores of the top-level key-value pairs found in the target and predicted YAMLs.
Similarly for playbooks the scores of its top-level key-value pairs are averaged, where the score of each of its tasks is computed as above.
The \textit{``name''} key and its value can be ignored as they has no effect on the execution of the task.
% and computes the average of the scores for the other key-value pairs in both the target and predicted task YAMLs.
The score for each key-value pair is the average of the key and value scores.
Currently keys missing from the prediction are given a score of 0, while keys inserted in the prediction are ignored.
%If a key is in both YAMLs it has a score of 1 which is averaged with the score of its value
If a key's value is a list or dict, its score is recursively computed by averaging the scores of each dict entries or list items.
When comparing the module names they are first replaced by their fully qualified collection name (FQCN) if necessary, e.g. \textit{copy} is changed to \textit{ansible.builtin.copy}.
Another normalization that is applied is to convert the old $k_1=v_1$, $k_2=v_2$ syntax for module parameters into a dict.
There are some modules that are almost equivalent, e.g. \textit{command} / \textit{shell}, \textit{copy} / \textit{template}, \textit{package} / \textit{apt}, \textit{dnf}, \textit{yum}.
Since they accept many of the same arguments and in some cases can be exchanged,
such module differences are given a partial key score which is averaged with the score of their arguments.

Our motivation for ignoring insertions 
%in these last two metrics 
is that they are less costly than deletions as they can be easily removed, but we plan to investigate the impact of including an insertion penalty.\\

\noindent\textbf{Schema Correct}: this metrics is designed to measure the correctness of the result, i.e. whether or not it satisfies the Ansible schema.
It does not reflect the accuracy of the model, as it applies just to the predictions.  The Ansible playbook and tasks schema used by the Ansible linter are quite strict and do not accept some historical forms which are still allowed by Ansible itself.  Hence a low score does not necessarily mean that the results would be rejected by Ansible. 
Since we did not filter our training data with these schema a sample with a perfect Exact Match score may have a Schema Correct score of 0.

\subsection{Results}
\begin{table*}
\centering
\begin{tabular}{p{0.18\textwidth}*{5}{|C{0.1\textwidth}}}
\hline
\multirow{2}{*}{\textbf{Model}} &
\multicolumn{5}{c}{\textbf{Dataset}} \\
\cline{2-6}
& The Pile & BigQuery & BigPython & Ansible YAML & Generic YAML \\
\hline
\hline
\CGNL{} & \checkmark & & & & \\
\CGMU{} & \checkmark & \checkmark & & & \\
\CGMO{} & \checkmark & \checkmark & \checkmark & & \\
\WA{} & & & & \checkmark &  \\
\WY{} & & & & \checkmark & \checkmark \\
\WAM{} & \checkmark & \checkmark & & \checkmark & \\
\WYM{} & \checkmark & \checkmark &  & \checkmark & \checkmark\\ \hline
\end{tabular}
\caption{\label{model-name-dataset} Model names and their associated pre-training datasets. The Pile, BigQuery and BigPython were used by Salesforce~\cite{nijkamp2022conversational}, while Ansible YAML and Generic YAML are introduced in this work.}
\label{model_table}
\end{table*}

\subsubsection{Pre-training}

\textbf{Pre-trained Models for Comparison} All the models are implemented with the same architecture as \CG{},  but are pre-trained on different datasets.
Table~\ref{model_table} introduces the names of the models and the datasets they were pre-trained on.  
The first three models correspond to the original \CG{} checkpoints released by Salesforce~\cite{nijkamp2022conversational}: \CGNL{}, \CGMU{}, and \CGMO{}. The last four rows are the domain-specific pre-trained \WIS{} models proposed in this paper for YAML data.
\WA{} was pre-trained only on Ansible YAML while \WY{} was pre-trained on both Ansible YAML
and Generic YAML. 
\WAM{} was initialized with the weights of \CGMU{} and we extended the pre-training
using Ansible YAML.   
\WYM{} was also initialized with the weights of \CGMU{} and we extended the pre-training
using both Ansible YAML, and Generic YAML.\\ 

\noindent\textbf{Experiment Settings} We first evaluate the models in few-shot setting on our Ansible test set which includes 
a distribution of the four generation tasks described previously: \PBNLtoT{}, \NLtoPB{},
 \TNLtoT{}, and \NLtoT{}.
Our main goal here is to understand how much adding Ansible and generic YAMLs to our pre-training
improves the performances of the models.   

Table~\ref{zeroshot_results} presents the results for all \CG{} and \WIS{} models as well as OpenAI Codex. 
For each row, we indicate the size of the model (i.e. number of parameters), as well as the size
of the inference context window. When the input to the model $\{Y_{NL},C\}$ (see~\ref{sec:prompt_formulation}) 
is larger than the context window, it is left-truncated.
For the tasks that do not include any context (\NLtoPB{} and \NLtoT{}), we found that adding the 
string ``\textit{Ansible\textbackslash n}'' prior to the prompt improved the performances of \CG{} models as 
well as Codex. For the \WIS{} models, we did not observe any significant change and therefore
left the context empty.       
All the models were evaluated using the four metrics described in Section~\ref{subsec:eval_metrics}:
Schema Correct, Exact Match (EM), \textsc{Bleu}, and Ansible Aware.  
In order to correctly evaluate against these metrics,
in the case of Ansible task generations, 
%(i.e. \PBNLtoT{}, \TNLtoT{}, and \NLtoT{}), 
we truncated the models output predictions to keep only the first generated task.   
For playbook generation (\NLtoPB{}), we did not apply any truncation. Finally, all results presented thereafter were obtained using greedy decoding. We would expect some improvement by using random sampling or beam search decoding.

\noindent\textbf{\CG{} Comparison on Ansible Generation}
The first three rows refer to the \CG{} models as released by Salesforce. As shown in Table~\ref{zeroshot_results}, \CGMU{} trained on The Pile and BigQuery performs the best among the three \CG{} models. Specifically,
\CGNL{} 350M performs the worse across all metrics with a \textsc{Bleu} of 24.95 and an Ansible Aware score
of 6.24. The Schema Correct is 71.26. This rather high value shows that the small subset of YAMLs present in the Pile is already
enough for the model to have a good understanding of the YAML syntax. However, note that this metric does not compare
against any target, and only indicates that \CGNL{} can generate correct Ansible YAML, $\sim$71\% of the time.  
\CGMU{} 350M scores are higher, especially the Ansible Aware score that improves by $\sim$28 points.
The improvement is mainly attributed to the very large amount of code present in the BigQuery ($\sim$120B training tokens).
The additional code samples help the model to have a better understanding of structures and syntax (e.g. indentation) 
as seen by the 12 point boost of Schema Correct.      
The results of \CGMO{} are similar to \CGMU{}, showing that the addition of more \textsc{Python} code 
does not help our Ansible generation tasks.
To measure the effect of the size of the model, we additionally compared \CGMU{} 350M, 2.7B, and 6B, as shown in Table~\ref{zeroshot_results}.
The larger models do perform slightly better, but the improvement is not striking. Comparing with the 350M baseline, 
the 2.7B model improves the Ansible Aware score by $\sim$1.8 points and the 6B model by $\sim$4.9
points.

\noindent\textbf{Codex for Ansible Generation}
We also evaluated Codex (\CODEXEDIT{}) on the Ansible generation tasks, as shown in Table~\ref{zeroshot_results}.
The Schema Correct and \textsc{Bleu} scores of Codex are in the same order of magnitude
as \CGMU{} 350M but the Ansible Aware is significantly higher (48.78).
Also note that the exact match is the highest of all models tested, which indicates
that Codex likely saw large portions of our Galaxy dataset.

%\todo{BL: Add an explanation for Codex having higher Ansible aware and lower schema correct.}

\noindent\textbf{\WIS{} models for Ansible Generation}
As shown in the last four rows in Table~\ref{zeroshot_results}. The \WIS{} models are notably better than \CG{} and Codex baselines, showing that our YAML pre-training
provided a boost in performance. 
The last two rows show the two \WIS{} models pre-trained with YAMLs only. 
Both models reach Ansible Aware score similar to Codex, and \textsc{Bleu} score
comparable to \CGMU{} 6B.  
\WAM{} 350M has the highest Ansible Aware score, $\sim$6 points higher than Codex
and $\sim$15 points higher than \CGMU{} 6B. The \textsc{Bleu} score is also $\sim$6 points better than \CGMU{} 6B.
These results show that adding a large collection of YAMLs to pre-train or extend the pre-training of an existing model
offer a large boost in performance for the Ansible task generations. Further, the \WIS{} models outperform \CG{} and Codex with less parameters, which is advantageous in this application that requires fast inference.

%Comparing the zero-shot results in Table~\ref{zeroshot_results} with the fine-tuned models in Table~\ref{FT_results}, 
%we observe a large improvement across all evaluation metrics. 
%Performance boost due to fine tuning is not surprising. However, we would like to quantify how much our 
% pre-training with YAML data help.

\subsubsection{Finetuning}
%RQ2
We fine-tuned and evaluated \CG{} and \WIS{} models on the Galaxy dataset described in~\ref{sssec:galaxy_dataset}.
As shown in Table~\ref{finetune_results}, fine-tuning on specified Ansible generation task is necessary and largely boost the performance compared to the few-shot results in Table~\ref{zeroshot_results}. For example, comparing \CGMU{} with 2048 context window in few shot vs. fine-tuned
\CGMU{} with the same context window, both \textsc{Bleu} Ansible aware scores increase by $\sim$30 points. To better understand how different experimental factors influence the Fine-tuned models, we conducted ablation studies regarding the format of prompt, pre-trained models, model size, context window size, and the dataset size.\\

\begin{table*}[htb]
\centering
%\begin{tabular}{p{0.31\textwidth}|*{5}{C{0.06\textwidth}}}
\begin{tabular}{p{0.2\textwidth}|*{6}{C{0.07\textwidth}}}
\hline
%\textbf{Model} & \textbf{Size} & \textbf{CW} & \textbf{BLEU} & \textbf{AC} & \textbf{AA}\\
\textbf{Model} & \textbf{\centering Size} & \textbf{\centering Context \\ Window}  & \textbf{\centering Schema \\ Correct} &  \textbf{\centering EM} & \textbf{\centering BLEU} & \textbf{\centering Ansible \\Aware}\\
\hline
\hline
\CGNL{} & 350M & 2048 & 71.26 & 1.69 & 24.95 & 6.24 \\
\CGMO{} & 350M & 2048 & 82.40 & 6.37 & 34.24 & 34.15 \\
\CGMU{} & 350M & 2048 & 83.65 & 6.92 & 34.26 & 34.40 \\
\CGMU{} & 2.7B & 2048 & 78.00 & 7.74 & 37.27  & 36.23 \\
\CGMU{} & 6B & 2048 & 85.80  & 7.98 & 39.67 & 39.27  \\
%\CGMU{} & 16B & 2048 & 73.60 & 7.56 & 35.49  & 35.98 \\
\hline 
%\textcolor{red}{\CODEXEDIT{}} & 12B & 2048 & 84.24 &  & 29.20  & 48.78 \\
\CODEXCOMPLETION{} & 175B & 2048 & 88.82 & 13.66 & 50.40  & 55.01 \\
\hline
\WAM{} & 350M & 1024 & 96.56 & 7.35 & 46.58  & 54.51 \\
\WYM{} & 350M & 1024 & 95.97 & 7.16 & 45.52  & 53.08 \\
\WA{} & 350M & 1024 & 95.10 & 4.63 & 39.49  & 48.03 \\
\WY{} & 350M & 1024 & 94.63 & 4.19 & 40.13  & 47.76 \\
\hline
\end{tabular}
\caption{\label{rq1} Evaluation results for \CG{}, Codex, and \WIS{} models in few-shot setting. The first section refers to the \CG{} models released by Salesforce, the second one to OpenAI Codex and the third one to \WIS{} models.
% While both models from the Codegen family and Codex have seen Ansible in their pre-training data, having a dedicated pre-training on such dataset improves performances over all the metrics. Starting from the Codegen checkpoints and then continue pre-training on 
% YAML files (\WAM{} and \WYM{}) gives the best results. Details about number of parameters and the input size the model can process are reported together with the metrics described in Section~\ref{subsec:eval_metrics}.
}
\label{zeroshot_results}
\end{table*}

\begin{table*}[htb]
\centering
%\begin{tabular}{p{0.31\textwidth}|*{5}{C{0.06\textwidth}}}
\begin{tabular}{p{0.2\textwidth}|*{6}{C{0.07\textwidth}}}
\hline
%\textbf{Model} & \textbf{Size} & \textbf{CW} & \textbf{BLEU} & \textbf{AC} & \textbf{AA}\\
\textbf{Model} & \textbf{\centering Size} & \textbf{\centering Context \\ Window} & \textbf{\centering Schema \\ Correct} & \textbf{\centering EM}  & \textbf{\centering BLEU}  & \textbf{\centering Ansible \\Aware}\\
\hline
\hline
\CGMU{} & 350M & 512 & 97.77  & 22.30 & 61.75 & 64.84 \\
\CGMU{} & 350M & 1024 & 98.06 & 28.64 & 66.03  & 69.77 \\
\CGMU{}  & 350M & 2048 & 98.02 & 27.14 & 66.12  & 69.69 \\
\CGMU{}  & 2.7B & 1024 & 98.36 & 28.03 & 65.25  & 69.41 \\
\hline 
\CGMU{}-prefix & 350M & 1024 & 72.96 & 12.37 & 56.29  & 45.87 \\
\hline
\WAM{} & 350M & 1024 & 98.00 & 29.36 & 66.67 & 70.79 \\
\WYM{} & 350M & 1024 & 98.02 & 28.79 & 65.92 & 69.65 \\
\WA{} & 350M & 1024  & 97.68 & 23.44 & 61.94 & 66.29 \\
\WY{} & 350M & 1024 & 97.97 & 23.27 & 61.20  & 65.70 \\
\hline
\WAM{} -50 & 350M & 1024 & 98.10 & 27.90 & 65.46 & 69.79 \\
\WAM{} -20 & 350M & 1024 & 98.08 & 25.00 & 63.37  & 67.90 \\ 
\WAM{} -10 & 350M & 1024 & 98.08 & 22.62 & 61.68  & 66.23  \\
\hline
\end{tabular}
\caption{\label{rq2} Evaluation results of the fine-tuned models. The first section shows results of \CGMU{} fined-tuned on Galaxy, varying the size of the context window and the number of parameters. The second section shows the \WIS{} models fine-tuned on Galaxy, for a fixed 1024 context window.
The last section corresponds to \WAM{} fine-tuned on Galaxy and varying the amount of data (10\%, 20\%, and 50\% of the dataset).
% the gain we have if, before fine-tuning on Ansible Galaxy, we extend the pre-train with more YAML files. The last section shows that if there is a round of pre-training before fine-tuning, we can achieve good performance without the full fine-tuning dataset. 
}
\label{finetune_results}
\end{table*}

\noindent\textbf{Effectiveness of Prompt Formulation}
As mentioned in Section~\ref{sec:prompt_formulation}, to take advantage of the feature of Ansible-YAML data, we re-formalized the code generation problem into a code completion process by utilizing the natural language prompt as a part of ``name'' field. To validate its effectiveness, we compare it with the typical prefix-based \CG{} model (named as \CG{}-prefix) which contains the prefix term``context code'' before the context information and ``prompt'' before the natural language part. According to the results shown in Table~\ref{finetune_results}, under the window size 1024, \CG{} with the proposed prompt format largely outperforms \CG{}-prefix, for example, ~10\% percent higher on \textsc{BLEU}, ~26\% percent higher on \textsc{Schema Correct}, and ~16\% percent higher on \textsc{EM}.  
\\

\noindent\textbf{Analysis on Different Pre-trained Models}
The pre-trained models play an important role in the performance of fine-tuned models. By comparing the fine-tuned \CGMU{} and \WAM{} (both with window size 1024 and model size 350M) in Table~\ref{finetune_results}, it shows pre-training on Ansible data can help the large language model better understand the syntax and structure of Ansible. Specifically, \WAM{} has gained 1\% increase regarding Schema Correct, EM and Ansible Aware.\\

\noindent\textbf{Analysis on Context Window Size}
We first compare fine-tuned \CGMU{} models with different context window sizes.  
More context improves the model predictions at inference time, but it also requires more compute resources for training.
In addition, in the case of our Ansible-specific generation tasks, this is not obvious whether a very large context
improves the model outputs.
In table~\ref{FT_results}, the first three rows present results for context window sizes of 512, 1024, and 2048, respectively.         
The 512 context window has a 61.75 \textsc{Bleu} and a 64.84 Ansible Aware score.
When doubling the size of the context to 1024, the \textsc{Bleu} goes up to 66 and the Ansible Aware score to 69.77.
However, we do not observe improvement when going beyond 1024,
as seen with the 2048 results.  
Note that this observation is based on our current dataset, and this is possible that other test sets
would benefit further from larger contexts. Nonetheless, in our current setup, we conclude that a 1024 context window 
is adequate. \\

\noindent\textbf{Analysis on Number of Training Data}
To investigate the influence of the training data, we use varying amount of data (10\%, 20\% and 50\% of training data) for finetuning. The comparison results are shown in the last three lines in Table~\ref{finetune_results}. With the increment of training data, the performance improved accordingly, for example, \textsc{BLUE} from 61.68\% to 66.67\%. However, the speed of improvement decreases, from 1.7\% per 10\% of data to 1.2\% per 50\% of data. This shows the current training data size has almost converged and the current fine-tuning data size is selected with the high performance cost ratio. 
It is interesting to note that by finetuning with even a little bit of data, \textbf{the Wisdom model performance on Ansible-YAML Generation task becomes much better than \CODEXEDIT~(in few-shot settings) on all metrics.}
As can be seen in Tables~\ref{zeroshot_results} and ~\ref{finetune_results}, the best performing Wisdom model, \WAM{}, trained on 100\% finetuning data is better than \CODEXEDIT{} in fewshot settings by about 15 BLEU points and about 16 EM points.

\begin{table*}[htb]
\centering
\begin{tabular}{p{0.12\textwidth}|*{5} {C{0.07\textwidth}}}
\hline
%\textbf{Generation Types} & \textbf{BLEU} & \textbf{AC} & \textbf{AA}\\
\textbf{Generation Types} & \textbf{\centering Count} & \textbf{\centering Schema \\ Correct} & \textbf{\centering EM} &  \textbf{\centering BLEU}  & \textbf{\centering Ansible \\Aware}\\
\hline
\hline
\textsc{ALL} & 50580 & 98.06 & 28.64 & 66.03  & 69.77  \\
\NLtoPB{} & 550 & 93.09 & 0.0 & 22.76  & 23.16 \\
\NLtoT{} & 6961 & 96.51 & 5.17 & 45.46 & 49.28 \\
\PBNLtoT{} & 3441 & 98.75 & 46.00 & 79.66 & 82.31 \\
\TNLtoT{} & 39628 & 98.35 & 31.65 & 69.41  & 72.93 \\
\hline
\end{tabular}
\caption{\label{rq5} Breakdown of the evaluation metrics per generation type (see~\ref{subsec:generationtypes}) for \CGMU{} fine-tuned on Galaxy. ``ALL'' refers to all generation types combined together.
}
\label{FT_results}
\end{table*}

\noindent\textbf{Analysis on Generation Types}
As discussed in Section~\ref{subsec:generationtypes}, there are 4 types of generation problems for Ansible-YAML Generation. To validate how the proposed model deal with these 4 types, we evaluate on them sepecreatly, as shown in Table~\ref{FT_results}. Due to the dominant number of \TNLtoT{} in the training data, the proposed model performs the best in this task. For \PBNLtoT{} type, even there are only 3441 fine-tuning samples, the performance is comparable to that of \TNLtoT{}. This may because \TNLtoT{} and \PBNLtoT{} are both used for generating a task given the natural languages and context ansible data and thus can benefit each other while fine-tuning. The proposed model has difficulty in generating a playbooks, as shown in the second line of the table, which is because of the limited number (i.e., 550 counts) of training data for \NLtoT{}. In addition, by comparing the performance between \PBNLtoT{} and \NLtoT{}, the necessities of utilizing the context information for generation is validated. Though \NLtoT{} has more training data than that of \PBNLtoT{}, the performance on \NLtoT{} decrease dramatically compared to that of \PBNLtoT{}, for example, 33\% decrement in \textsc{BLEU} .
\section{Conclusion}
This work describes the application of transformer-based models to the generation of Ansible-YAML, starting from a user-provided natural language prompt. The objective of this model is to build an AI assistant for Ansible users and improve their productivity. We provide a formal definition of the problem and we start from an existing pre-trained decoder-based model. We built a new training dataset with Ansible data for code generation that will be shared with the community. We extend the training of the base model with our dataset and evaluate the results. Our results show that with our approach, Wisdom performs Ansible generation equally or better compared to state-of-the-art models for code generation.

\section*{Limitations}
A lot of Ansible development happens on playbooks. 
However, playbooks are not well represented in our fine-tuning dataset since we found very few acceptable playbook samples in Ansible Galaxy. 
And most of the ones that are included are small with two or less tasks.
Ansible Blocks, which are logical groups of tasks are also something we have not specifically trained and tested on.
This is something we hope to expand to in the future.

We also hope to do more analysis on the models sensitivity to prompts and robustness to changes in indentation, quotes and letter case.
Currently we focus on the Natural Language to Ansible generation task. 
This can be expanded to a more general completion task where a user can prompt the model at any stage of code development.

\section*{Ethics Statement}

\subsection{Legal Implications}
Wisdom is trained on code repositories that are publicly available and with an open-source license. Training of ML algorithms on public repositories, such as those in GitHub, has been regarded has fair use~\cite{trainingIP}.
Once trained, even if it is a rare occurrence, Wisdom could potentially generate code that is identical to a training set sample. When this happens, the generated code is most likely a very common pattern in Ansible, rather than the result of a copy. Furthermore, while Wisdom provides a recommendation, it is the user’s choice to accept it and use it in the codebase.

\subsection{Offensive Language}
Large amounts of public repositories could contain language that is offensive or discriminatory to multiple groups in the form of comments or code. While this is primarily a research work, not intended for product use, a product ready version of the model would undergo a major data cleaning and normalization process to avoid the generation of unwanted expressions.

\subsection{Security and Safety Risks}
Wisdom is trained on good quality data, however there is a significant risk of generating Ansible that contains security vulnerabilities or could damage a system. The model it is not trained to generate Ansible that is secure or safe, but only to optimize metrics with respect to our test set. While security vulnerabilities cannot be completely eliminated, it is possible to reduce this event by explicitly improving the security and safety of training data, and also by performing basic post-processing analysis to avoid the most common vulnerabilities. Both approaches would be considered in a product-ready version of the model. 

\subsection{Economic and Labor Market Impact}
The topic of economic and labor market disruption by AI algorithms has been the subject of a wide range of arguments. Specifically, an AI coding assistant could potentially be seen as a threat to software development labor. A deeper look at how these coding assistants should and are being used ~\cite{chen2021evaluating} will clearly highlight how the presence of an human expert cannot be replaced. Indeed, current ML models cannot provide the deep semantic comprehension needed to understand and integrate the recommendation into the codebase.

\bibliography{custom, anthology}
\bibliographystyle{IEEEtran}

%\bibliographystyle{./IEEEtran}
%\bibliography{./custom,./anthology}

%\appendix
%\label{sec:appendix}
%\input{sections/appendix}

\end{document}